\documentclass[10pt, conference, compsocconf]{IEEEtran}
\usepackage[pdftex]{graphicx}
\usepackage[cmex10]{amsmath}

\usepackage[tight,footnotesize]{subfigure}
\usepackage{url}
\usepackage[ruled,vlined,linesnumbered]{algorithm2e}
\usepackage{lipsum}

\usepackage{flushend}

\let\OLDthebibliography\thebibliography
\renewcommand\thebibliography[1]{
  \OLDthebibliography{#1}
  \setlength{\parskip}{0pt}
  \setlength{\itemsep}{0pt plus 0.3ex}
}

\newcommand{\remove}[1]{}

\begin{document}

\title{CloudTree: A Library to Extend Cloud Services for Trees}

\author{\IEEEauthorblockN{Yun Tian\IEEEauthorrefmark{1},
Bojian Xu\IEEEauthorrefmark{1},
Yanqing Ji\IEEEauthorrefmark{2},
Jesse Scholer\IEEEauthorrefmark{1}}
\IEEEauthorblockA{\IEEEauthorrefmark{1}Department of Computer Science,
Eastern Washington University, Cheney, WA 99004, USA}
\IEEEauthorblockA{\IEEEauthorrefmark{2} Department of Electrical and
  Computer Engineering, Gonzaga University, Spokane, WA 99258, USA}
\IEEEauthorblockA{\tt \{ytian,bojianxu\}@ewu.edu, ji@gonzaga.edu, jscholer@ewu.edu}
} 

\maketitle

\begin{abstract}
  % The abstract goes here. DO NOT USE SPECIAL CHARACTERS, SYMBOLS, OR
  % MATH IN YOUR TITLE OR ABSTRACT.
  % motivation%what is about%what we did%the findings
  % Along comes with big data management is the metadata organization.
  In this work, we propose a library that enables on a cloud the creation and
  management of tree data structures from a cloud client. As a proof
  of concept, we implement a new cloud service \emph{CloudTree}.  With
  CloudTree, users are able to organize big data into tree data
  structures of their choice that are physically stored in a cloud.
  % Built upon
  % existing cloud APIs, CloudTree is a client-side library that
  % enables
  % users to create, query, and update tree data structures that are
  % physically stored in a cloud.
  % 
  % We describe the motivation, design, and implementation of
  % CloudTree.
  We use caching, prefetching, and aggregation techniques in the
  design and implementation of CloudTree to enhance performance.
  We have implemented the services of Binary Search Trees (BST) and
  Prefix Trees as current members in CloudTree and have benchmarked
  their performance using the Amazon Cloud. The idea and techniques in
  the design and implementation of a BST and prefix tree is generic and
  thus can also be used for other types of trees such as B-tree, and
  other link-based data structures such as linked lists and graphs. 
  Experimental results show that CloudTree is useful and
  efficient for various big data applications.
   %including those that demand high performance.

\end{abstract}


% For peer review papers, you can put extra information on the cover
% page as needed:
% \ifCLASSOPTIONpeerreview
% \begin{center} \bfseries EDICS Category: 3-BBND \end{center}
% \fi
%
% For peerreview papers, this IEEEtran command inserts a page break and
% creates the second title. It will be ignored for other modes.
\IEEEpeerreviewmaketitle

\section{Introduction}
\label{sec:intro}
Due to the ever-increasing data size caused by advances in
electronic and computational technologies, computer scientists have
been making efforts in finding efficient solutions to challenged involving 
large-scale data. Such efforts in this direction include external memory
computing~\cite{AV1987,vitter:IO-book}, cache-oblivious
computing~\cite{FLPR1999}, succinct and compressed data
structures~\cite{NM2007}, and distributed and parallel computing~\cite{AW2004,GKKG2004}.

Cloud computing proposed in late 2000s is an idea extended from
distributed computing. It centralizes the computing resources and
their management into one place (logically) and provides the usage of
the resources as ``utility services''. By doing so, users are freed
from complexities in hardware infrastructure management and 
only pay for the exact cost of the amount of resources that they
need. Since the hardware infrastructure of the cloud is hidden from the 
user, the data storage capacity and computing power provided by the
cloud is conceptually unlimited, which is critically important in the
era of big data.

On the other hand, because users do not have control of the
low-level hardware infrastructure, it becomes inefficient or
impossible for users to perform certain computations. For example, if one
wants to create and use an indexing tree data structure for a large
data set, the current solution that is available for the user is to
rent a physical or virtual machine (VM) from the cloud, then treat the VM as their
own local machine and do the computation as needed on the VM, which
conceptually provides unlimited storage capacity and computing power
as long as the user continues to pay for the rental. However, there are two critical
disadvantages of this solution: 1) The rental of a VM is often more
expensive than the mere data storage space rental, provided the same
amount of storage space is rented.  2) The user has to pay the rental
of the VM, even if no computation work is being conducted, because
once the lease of the VM is over, all the data stored within the VM
may be permanently lost. 

%3) In some scenarios, users may not want to
%ship the computation of their applications to the cloud for reasons of
%security or money saving (because their local machine's computing
%power is good enough). Instead, users only want to treat the cloud as
%their extended memory and use the cloud to save their data structures
%that serve the computation on their local machine. 

Trees are important linked data structures.
%which are crucial in data indexing for fast data retrieval. 
 When data size is large, often it is necessary to organize the dataset 
into a tree structure to avoid a linear scan. For example, 
we may frequently search a set of common prefix strings in a large group of target strings, where
a prefix tree is quite useful. 
In another scenario, in fields of graphics and particle physics, numerous objects 
with three dimensional coordinates are queried and processed on a regular basis, 
in which a k-d tree or an R-tree is quite useful to organize these spatial data objects. 
Again, existing cloud storage services that provide underlying hashing, sorting or search tree indices 
are not effective for these \emph{spatial} data items. %that consist of multi-dimensional components. 
In these cases, users have to build their own trees in a cloud.
%when data is intended to be stored in cloud.

In this work, we propose a library that enables a 
new cloud service \emph{CloudTree}.  With CloudTree, users are able
create, manage, and use tree data structures for big data sets on a
cloud without using VM. Users will perform the computation of
their application on their local machine and treat the cloud as
their extended memory, where the tree is being stored. 
We make the following contributions.  

%We describe our motivation, design and
%implementation of the CloudTree in this paper. In addition, 
%we benchmarked the performance of CloudTree using the Amazon Cloud.
%Experiment results suggest CloudTree service is promising and can be used 
%in various big data applications, even in performance-demanding applications. 

\smallskip 
\noindent
1) To our best knowledge, there is no existing cloud storage service
that directly allow users to create, manage, and use an
\emph{external} tree on cloud.  We are first to propose this concept
and implement it as a cloud client-side library.

\noindent
2) 
Because of the scalability and unlimited storage space offered
by cloud, CloudTree enables users to create trees
of \emph{unlimited} size in their applications.  In this case,
users need not be concerned about the limited size of their local RAM.

\noindent
%3) Since CloudTree only uses the cloud for storage, the cost of
%  using CloudTree will be much cheaper than renting a VM, provided
%  the same amount of storage space is rented. 

\noindent
3)
We adopt several optimization techniques to address the challenge of expensive network communications
between an user's local computer and a cloud. We utilize
caching, prefetching, and aggregated operations, which significantly
improve the performance of CloudTree.  We implement and benchmark the
performance of CloudTree using the Amazon Cloud.
Experimental results suggest CloudTree is promising and can be
quite useful in various big data applications, including
performance-demanding applications.

\noindent
4) The idea of CloudTree can be easily used for other 
  link-based data structures, such as linked list and graphs.

\section{Related Work}
\label{related}
%We focus the survey of related work on the particular subfield of 
%managing and using data structures on a cloud. 
Much research has been focused on utilizing tree-based structures in
the cloud for the purposes of indexing. One such work involves the use
of an A-tree \cite{Papadopoulos11}: a distributed architecture that
combines bloom filters and R-trees for fast index queries. The work in
\cite{Dehne13} also utilizes the cloud to store a distributed tree in
an attempt to achieve real-time on-line analytical processing. Similar
to CloudTree, the work in~\cite{Yu14} uses a B-tree contained in a
NoSQL database (MongoDB) for indexing. However, none of these works
allow for users to manipulate the data, nor is the tree generic. Each
utilize specific tree structures whose main purpose is for indexing,
not for general use.

A number of related works allow for the creation of tree structures on
Amazon's EC2 service. The solution proposed in \cite{Wu10} is that of
a dynamic B-tree indexing scheme that resides in the cloud. HSQL
\cite{Chang12} similarly utilizes NoSQL databases to store a
distributed B-tree. In \cite{Mo14}, a tree-based structure is proposed
that ensures database delete operations are external and
permanent. One major drawback to these works is their reliance upon
Amazon EC2; because the structure is implemented on a virtual machine
in main memory, it is not accessible from outside the cloud. CloudTree
allows for external storage of any generic tree structure that is
permanent without the need for a constantly running virtual machine.

Rather than focusing on cloud tree storage, much research is devoted
to indexing large amounts of data in a tree structure stored on
external memory. B and B+ trees~\cite{BM1970} have had a long history of
success in indexing large data sets on external memory.  String
B-tree~\cite{ferragina:string_B-tree} and its cache-oblivious
version~\cite{bender:oblivious-string-B-tree} serve the indexing of
large strings for pattern matching on external memory.  Behm et
al.~\cite{Behm11} proposes a combination of tree and index list to
quickly query indexes stored in external memory to balance I/O and
query time. This related research attempts to confront similar
challenges in storing tree structures on external memory that allow
for fast creation, querying, and updating. However, these solutions
lack the scalability that CloudTree provides, as well as integration
with cloud services.

%\begin{figure}[t]
%\begin{center}
%\includegraphics[scale=0.45]{pictures/intersect.pdf}
%\caption{Intersected Objects in 2D. The dotted square represents the spatial query, 
%and intersected objects are shown as shaded rectangles.}
%\label{fig:intersect}
%\end{center}
%%\vspace{-6 mm}  %originally not here
%\end{figure}
%%\vspace{-3 mm}  %original no here

\section{Design of a CloudTree}
\label{design}
In this section, we describe general concepts about CloudTree,
including API design, CloudTree node representation, generic design and
advantages of CloudTree. 

\begin{figure}
\begin{center}
{\tiny \tt 
\begin{tabular}{rl}
1 & import ewu.ytian.CloudTree;\\
2 & $\ldots \ldots$\\
3 & // Note CloudTree is the parent class of CloudPrefixTree and
CloudBSTree\\
4 & CloudTree ctrie = new CloudPrefixTree(treeName1, newTreeOrNot);\\%//a prefix tree\\
5 & CloudTree cbst = new CloudBSTree(treeName2, newTreeOrNot);\\%//a BST\\
6 & ctrie.init();\\
7 & cbst.init();\\
8 & ctrie.insert(``phrase1'');\\
9 & cbst.insert(num1);\\
10 & $\ldots \ldots$\\
11 & ctrie.query(``phrase1''); //return true or false\\
12 & cbst.delete(num3);\\
13 & $\ldots \ldots$\\
14 & ctrie.delete(``phrase3''); //delete phrase3 from the tree ctrie.\\
15 & cbst.query(num1); //return true if num1 exists in the tree
cbst.\\
16 & $\ldots \ldots$\\
17 & ctrie.close(); //paired with init() call \\
18 & cbst.close(); //paired with init() call\\
\end{tabular}
}
\caption{Sample code using CloudTree APIs to create and manipulate trees in a cloud. }
\label{fig:usecase}
\end{center}
%\vspace{-6 mm}  %originally not here
\end{figure}

%\vspace{-3 mm}  %original no here

\subsection{API Design}
%demo how to use the service in a general way.
%cloudTree ctree = new CloudTree();
%ctree.init();
%.....
%ctree.close();

\begin{figure}
\begin{center}
 \begin{minipage}[c]{1.2in}
{\tt \scriptsize
\begin{tabular}{|l|}
 \hline
node\_id:v0\\
\hline
data\_item:v1\\
children\_set:\{\\
\ \ (child\_id1:v2), \\
\ \ (child\_id1:v3), \\
\ \ \;$\ldots\ldots$\\
\}\\
otherAttrib:v4\\
\hline 
\end{tabular}
}
\end{minipage}
 \begin{minipage}[c]{1.0in}
\caption{The design of a CloudTree node,  where each tree node is identified by a numerical tree node id. }
\label{fig:nodedesign}
\end{minipage}
\end{center}
\end{figure}

Figure \ref{fig:usecase} shows a piece of sample code that calls the CloudTree APIs.
At lines 4 and 5, we create a cloud prefix tree instance \emph{ctrie} and a cloud binary search tree instance \emph{cbst} respectively.
We provide a string $treeName_i$ to uniquely identify each tree in cloud. We can also specify whether the tree instance will be initialized using an existing tree in cloud by providing a second parameter \emph{newTreeOrNot}. %Inconsistent information users provided results in an error.
%For example, it is an error to instantiate a CloudTree instance with a tree name \emph{T} and \emph{newTreeOrNot = false}, when the tree name \emph{T} is not found in the system.

In figure \ref{fig:usecase}, the init() method performs authentication and sets up parameters used to communicate with the cloud.
Then, users can use the CloudTree instance as if it is stored in local RAM, to insert, query and delete data items in the tree.
After finishing all operations, the user is required to call the close() method to clean up resources and to permanently save the tree name and other information.
%Metadata for a CloudTree is discussed in the section \ref{metadata}.

\subsection{CloudTree Node Representation}
Each CloudTree node is modeled as an object, similar to a JSON object.
Figure \ref{fig:nodedesign} shows the general design for a CloudTree node.
We use a collection of attribute-value pairs to describe a tree node. For example, in the 
top box of figure \ref{fig:nodedesign},
we identify each node with a pair $(node\_id : v_0$), where an attribute named \emph{node\_id} --- 
together with a numerical value $v_0$ --- are used as a reference 
to uniquely determine a tree node.
In addition, if needed, each node has an attribute \emph{data\_item} to describe a
data item (or a key) stored in a node.

Another important attribute in a CloudTree node is \emph{children\_set} which records a set of 
child references and, if needed, the data item stored in each children. The child set is shown in figure \ref{fig:nodedesign} in a format 
$\{(child\_id_1 : v_1), (child\_id_2 : v_2), ....\}$, where $child\_id_i$ refers to a numerical child tree node id, and $v_i$ denotes the data item
stored in that child. 
%It is worth mentioning that because the children are represented with a 

It is quite flexible to add other attributes in a CloudTree node as needed.
For example, an attribute \emph{leaf} describes whether the node is a leaf node or not.
In another example, for a prefix tree, we need an attribute \emph{word} to 
indicate whether the path from the root to the current node constitutes a valid English word
in a dictionary.

\subsection{Generic Design}
 \begin{figure}[t]
\centering
\subfigure[A generic tree structure.]{
  \includegraphics[scale=0.4]{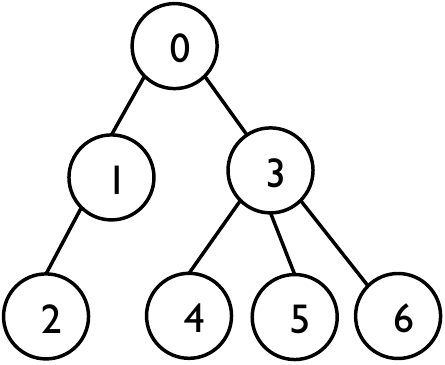}
   \label{fig:gtree}
   }
\hspace*{1cm}
\subfigure[CloudTree representation.]{
  \includegraphics[scale=0.45]{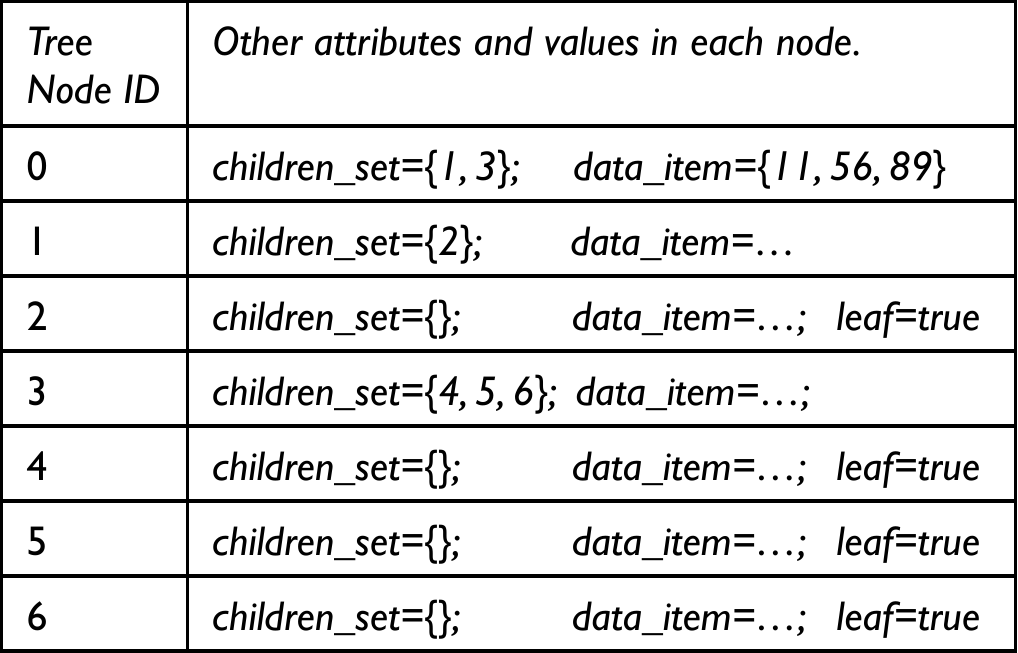}
   \label{fig:gstore}
   }
 \caption[]{A generic tree is shown in figure(a). Figure(b) shows its CloudTree representation.
}
 %\vspace{-4 mm}  %originally not here
 \label{fig:generic}
\end{figure}

In figure \ref{fig:generic}, we describe a simple tree structure and its representation using CloudTree.
The number in each node in figure \subref{fig:gtree} is a node id. 
Traditional child references (or child links) in each node are represented by a set of child node ids. 
For example, node $3$ has three children with ids \{4, 5, 6\}. 
Therefore, we can store a CloudTree instance either in a file or in a database table in a cloud,
with each tree node considered as an item in the file or in the table. Then,
we use the unique tree node id as a primary key to retrieve a tree node. 
It is worth mentioning that the design can be applied to other linked structures as well, for example, 
linked lists or graphs.

\begin{figure}[t]
\begin{center}
\includegraphics[scale=0.45]{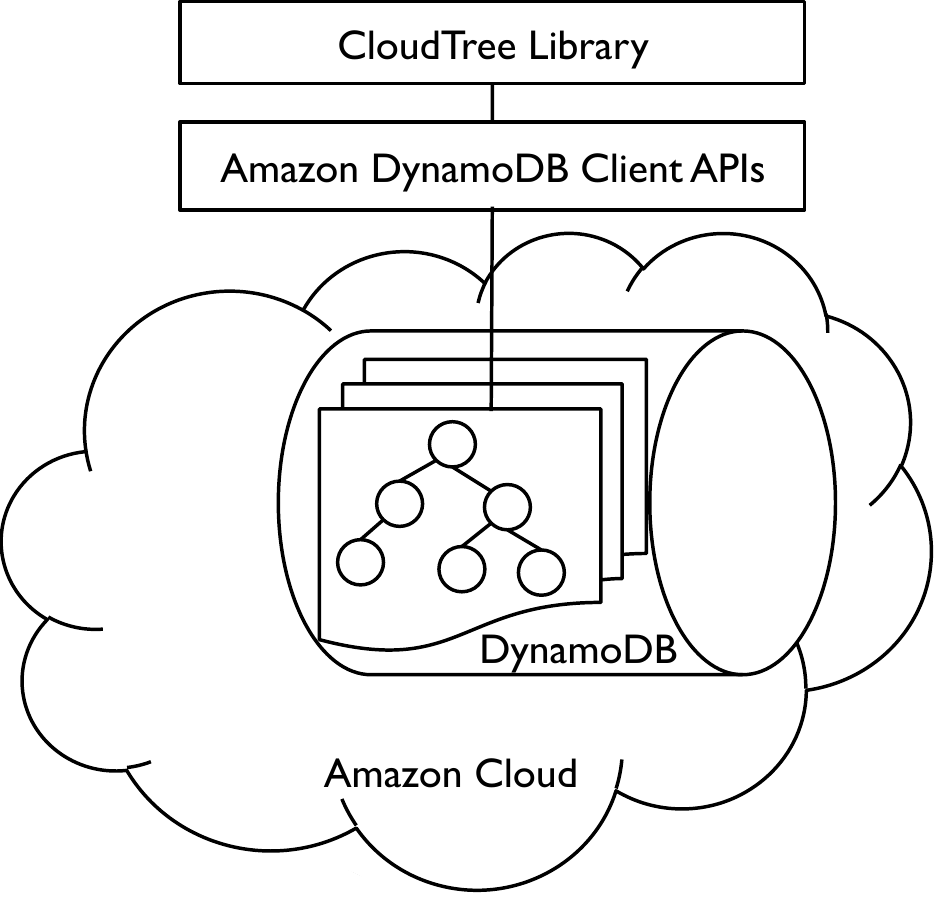}
\caption{Overview of the CloudTree Library Implementation. A CloudTree instance is stored in 
an Amazon DynamoDB database table, and is manipulated by using Amazon DynamoDB client APIs. }
\label{fig:impPic}
\end{center}
%\vspace{-6 mm}  %originally not here
\end{figure}
%\vspace{-3 mm}  %original no here

%%
%%
\subsection{Advantages of CloudTree}
%We first summarize the nice properties of a CloudTree in this section.
\begin{itemize}
%\centering
\item \emph{Generic}: First, the design of the proposed CloudTree is generic, 
allowing it to be applied to any existing tree data structure, such as prefix tree, B-tree, R-tree, K-d tree, etc.  
%are all able to fit in the design of CloudTree. 
Second, the CloudTree library can be implemented using cloud services from various cloud providers, such as Amazon, RackSpace or other providers.
\item \emph{External and Permanent}: A CloudTree instance is not stored in the traditional \emph{main memory (RAM)}, but in durable cloud storage instead.
Note that CloudTree instances are not stored in the RAM of a VM that is provisioned in a cloud either, such as an EC2 instance of Amazon.
%Data in CloudTree is permanent and durable. 
\item \emph{Unlimited Size}: Due to scalable cloud storage space, CloudTree's use of the underlying cloud storage results in a conceptually unlimited size.
\item \emph{As A Cloud Service}: With the proposed CloudTree library, authorized users can connect to a cloud and create a tree data structure in the cloud, anywhere that has Internet access.
\item \emph{Dynamic}: CloudTree supports tree creation, query, update and deletion operations. After a CloudTree instance is closed, the user may \emph{re-open} the existing tree again.
\item \emph{Transparent}: Users manipulate a CloudTree instance in the same way as if the tree
is stored in the RAM of a local computer. %The implementation details of CloudTree are encapsulated and hidden from the end user. 
\end{itemize}

 \begin{figure*}[t]
\centering
\subfigure[An example prefix tree represented using CloudTree, called CloudPrefixTree,
 after a set of English phrases have been inserted in an order \{``an", ``at",
	``and", ``bee", ``boy", ``hello", ``he"\}.]{
  \includegraphics[scale=0.51]{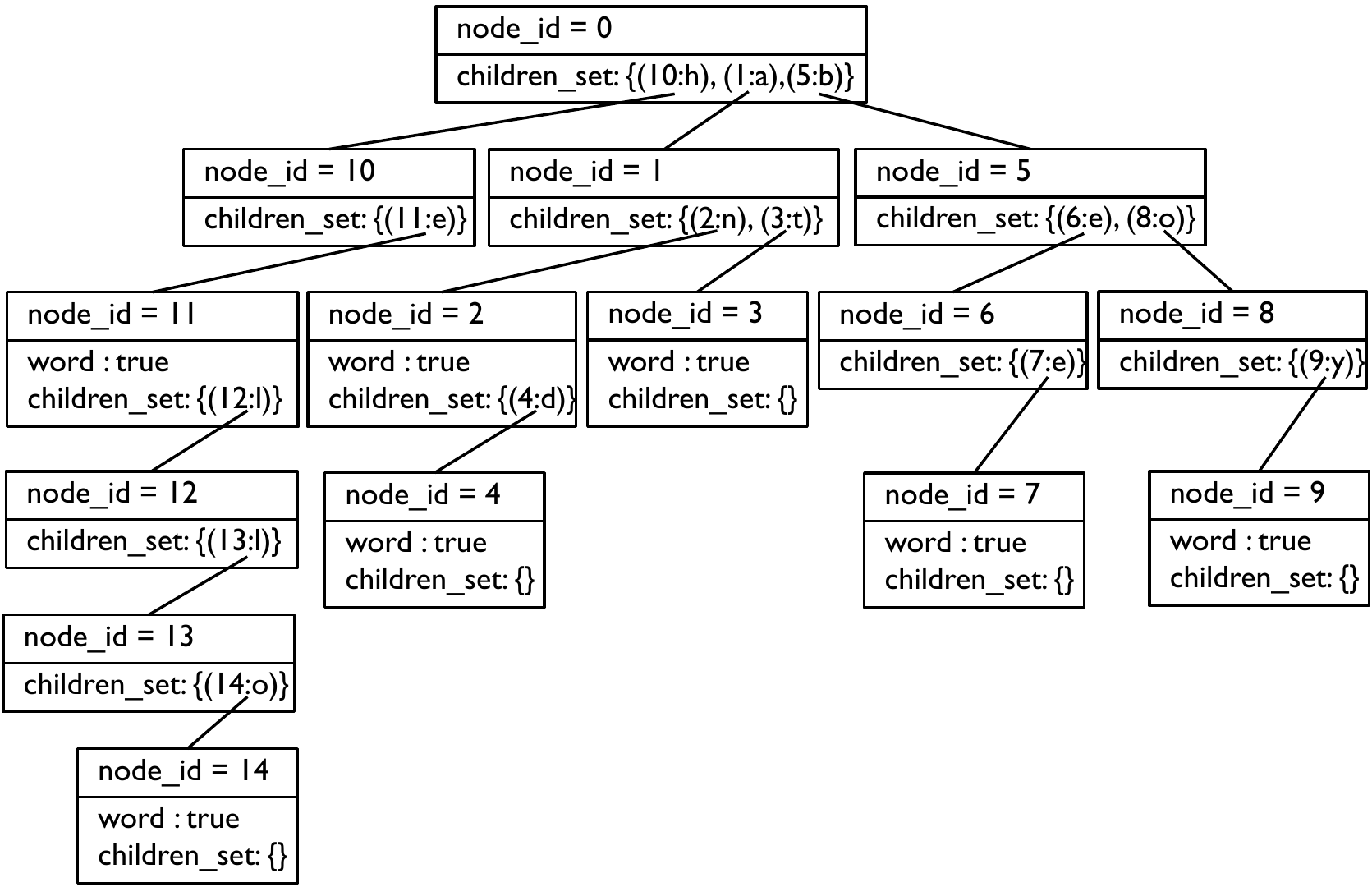}
   \label{fig:trieSample}
   }
\hspace*{1cm}
\subfigure[An example binary search tree represented using CloudTree, called CloudBSTree, after a set of integers have been inserted in an order \{5, 2,
	1, 3, 4, 7, 6, 8\}.]{
  \includegraphics[scale=0.51]{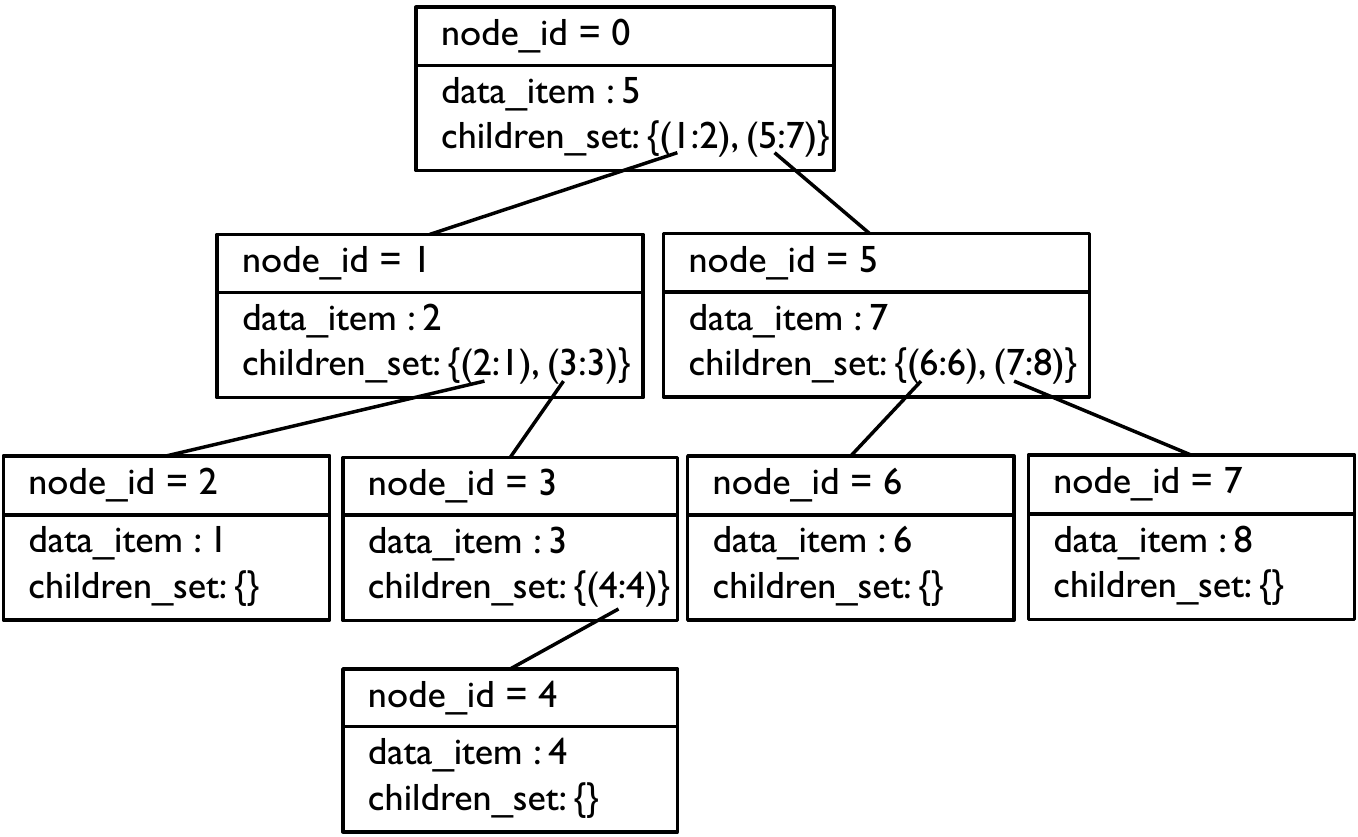}
   \label{fig:bstSample}
   }
% \subfigure[MAQR-tree node representations for 2D replicas in figure \subref{fig:newtree}.]{
%  \includegraphics[scale=0.65]{pictures/newgtr_3.pdf}
%   \label{fig:newgtr}
%   }
%   \label{fig:GTRtree3}
 \caption[Optional caption for list of figures]{Two instances of CloudTree: a CloudPrefixTree and a CloudBSTree.
node\_id  always increases when new elements are inserted into
a CloudTree. 
 %a new tree node\_id is generated by adding one to the current maximal node\_id value in the existing tree.
}
 %\vspace{-4 mm}  %originally not here
 \label{fig:examples}
\end{figure*}

\section{Implementation}
In this section, we describe an overview of implementation, 
the underlying storage and two 
members of CloudTree: CloudPrefixTree and CloudBSTree.

\subsection{Overview}
%why use dynamoDB

We use Amazon DynamoDB service to implement CloudTree.  
%Inparticular, we use Amazon DynamoDB Java client APIs to manipulate a
%tree data structure that is stored in Amazon DynamoDB database
%table. 
Figure \ref{fig:impPic} illustrates the implementation of
CloudTree.  The CloudTree library extends the underlying Amazon DynamoDB
client APIs to enable tree creation, insertion, query and deletion
operations in the cloud.  Each tree instance is stored in a separate
Amazon DynamoDB table.

There are three reasons why we chose Amazon DynamoDB among other
services.  First, access to data in
Amazon DynamoDB is fast with low read and write latency,   
%The server-side can respond to a request in single-digit milliseconds,
%partially because the data in DynamoDB is stored on Solid State
%Storage(SSD) devices
which
is quite suitable for performance-demanding
applications, such as gaming and shopping cart
services  \cite{AmazonService13}.  
Second, DynamoDB storage is scalable and durable with high
availability, two very desirable features that
are required by the CloudTree.  Third,
%Amazon DynamoDB is a NoSQL database service, 
%in which tables do not have fixed a schema, 
%so each data item can have a different number of attributes.
our CloudTree design nicely fits in the data model that DynamoDB provides,
which is discussed in the next section. 

%Reasons why amazon dynamodb includes: high performance low latency, stored on SSD.
%support real-time operations, we have two references here.
%
%Flexible NoSQL data model that fit in our CloudTree design.

\subsection{Data Model of DynamoDB} 
%Intro and data model of dynamodb.
%In Amazon DynamoDB, a database is a collection of tables. A table is a collection of items and each item is a collection of attributes.
%DynamoDB is a NoSQL database: Except for the required primary key, a DynamoDB table is schema-less. Individual items in a DynamoDB table can have any number of attributes,
%Each attribute in an item is a name-value pair. An attribute can be single-valued or multi-valued set.
Amazon DynamoDB is a NoSQL database. Except for the required primary key, a table in DynamoDB does not have a fixed schema. 
An item or object in such a table could have arbitrary number of attributes, with each attribute represented by a name-value pair.
Note that the value associated with an attribute could be a single value or a muli-valued set. 
These features in DynamoDB provide the means to represent an object that is traditionally stored in the main memory.
In this work, we consider each CloudTree node as an item in an DynamoDB table.
%Each Dynamo table is the storage media for an individual CloudTree instance. 
The DynamoDB provides client APIs
to insert, update or delete an item in a DynamoDB table.

\subsection{Cloud Prefix Tree Implementation}
\label{prefixtree}
%pseudo code of tree operations

\remove{
\begin{algorithm}[t]
{\small
  \caption{$query(String \ S)$: Query whether a prefix string $S$ exits in a CloudPrefixTree. If found, return true, otherwise return false.}
\label{algo:triequery}

\KwIn{The query string $S$.} 
%\KwOut{All LR's that cover position $k$ or find no such LR.}

\smallskip 

\tcc{Start the search with root node, with id = $0$.}
$curNodeID \leftarrow 0$\;

\smallskip

%\tcc{Calculate $lr_1,lr_2,\ldots,lr_n$.}
\For{$i=0, 1,\ldots,S.length - 1$}{
%$curChar \leftarrow S_i$\; %\tcp*{$\langle start, end\rangle$: start and ending position of $lr_k$.}

\tcc{Whether in current node there is a child associated with character $S_i$.}
$childNodeID \leftarrow hasChild(curNodeID, S_i)$\;

  \If(\tcp*[f]{Such a child for $S_i$ exists.}){$childNodeID\geq0$}{$curNodeID \leftarrow childNodeID$\;}
  \Else
   {$return \  false$\;}
}
$return \ true$\;
}
\end{algorithm}
}

\begin{algorithm}[t]
{\small
  \caption{$insert(String \ S)$: insert a prefix string $S$ into a CloudPrefixTree.}
\label{algo:trieinsert}

\KwIn{The string to insert $S$.} 
%\KwOut{All LR's that cover position $k$ or find no such LR.}

\smallskip 

\tcc{Start with root node, with id = $0$.}
\If(\tcp*[f]{root node not exist}){$nextNodeID = 0$}{$Create\ the\ root\ node\ with\ id\ =\ 0\ and\ insert\ into\ cloud\ DynamoDB\ table;$}

$curNodeID \leftarrow 0$\;

\smallskip

%\tcc{Calculate $lr_1,lr_2,\ldots,lr_n$.}
\For{$i=0,1,\ldots,S.length - 1$}{
%$curChar \leftarrow S_i$\; %\tcp*{$\langle start, end\rangle$: start and ending position of $lr_k$.}

\tcc{Whether in current node there is a child associated with $S_i$}
$childNodeID \leftarrow hasChild(curNodeID, S_i)$\;

  \If(\tcp*[f]{Such a child for $S_i$ exists.}){$childNodeID\geq0$}{$curNodeID \leftarrow childNodeID$\;}
  \Else
   { \tcc{Create a child string. Child id and the character $S_i$ associated with the child are delimited with  $:$, then add it to current node.}
   $child \leftarrow ``nextNodeID : S_i$''\;
   $addChild(curNodeID, child)\; $\tcc{Update tree node in cloud via internet.}
   break;
   }
}
\tcc{For each character that has not yet scanned in $S$}
\For{$j=i+1,i+2,\ldots,S.length - 1$}{
$Create\ a\ tree\ node\ N for\ S_j, with\ id = nextNodeID\ and\ with\ child\ id\ =\ nextNodeID + 1$\;
\smallskip
$Insert\ N\ into\ the\ cloud\ DynamoDB\ table$\;
$nextNodeID \leftarrow nextNodeID + 1$\;
}
\If(\tcp*[f]{If S is not a prefix of an existing string in the tree.}){$j == S.length$}{$Insert\ a\ leaf\ node\ indicating\ the\ end\ of\ S $\; }
}
\end{algorithm}

Figure \ref{fig:trieSample} presents a CloudPrefixTree instance, 
a prefix tree represented with the CloudTree design.
%Each tree node is identified by an unique numerical node id, shown in the top box of each node.
%The tree node id in the top box of each node is used as a primary key in the DynamoDB table, 
The tree node id is shown in the top box of each node,
on which a hash index has been constructed for fast data retrieval.
With provided DynamoDB client APIs,
we can retrieve a CloudTree node, given its tree node id.

We explicitly store child node ids for an node using the attribute $children\_set$, 
a set data structure with each element represented by a string.
Each element in the $children\_set$ contains two pieces of information,
 a child node id and the character
that is associated with that child. The two pieces of information share a single string, but 
is delimited by the special character `$:$' (colon).
For example, there are three children in the root node ($node\_id = 0$) in figure \ref{fig:trieSample},
with child node ids \{10, 1, 5\}. The corresponding characters associated with
these children are \{`h',`a',`b'\}.
Note that since we use a string set in DynamoDB for the attribute $children\_set$, 
string elements in the set maintain no order.

\remove{
Algorithm \ref{algo:triequery} shows the pseudo code for the CloudPrefixTree query operation.
For each character  $S_i$ in the query string, we search in the current tree node whether there exists a child that is associated with
 $S_i$. The function $hasChild()$ returns a positive $childNodeID$ for such a child if exists, otherwise returns a negative number.
 If the child does exist, that child becomes the current tree node and we repeat the previous step for the next character $S_{i+1}$.
 The $hasChild()$ function  has to retrieve the current tree node stored in the cloud via Internet. 
 }
 
We describe the CloudPrefixTree insertion operation in Algorithm \ref{algo:trieinsert}.
The variable $nextNodeID$ denotes the next available number for a new tree node id, which 
is increased by one after each tree node insertion.
Pseudo code lines $1$ and $2$ handle the edge case of root node insertion. 
If nodes already exist in the tree,
we start at the root node with tree node $id = 0$ at line $3$. In line $4$ through $11$, 
we search whether a prefix string already exits in the cloud tree that 
matches any prefixes \{$S[0], S[0, 1], \ldots, S[0,1,\ldots,i]$\} of string $S$.
The for loop breaks once the prefix with the maximum length has been found.
 At line $5$, the function $hasChild()$ searches in the current tree node whether a child that is associated with $S_i$ exists.
 It returns a positive $childNodeID$ if such a child exists, otherwise it returns a negative number.
The function $addChild(curNodeID,child)$ at line $10$ adds a child specified by a string parameter $child$ into the tree node with 
$id = curNodeID$.
We then create new nodes and insert the remaining unscanned characters $S[i+1, i+2, \ldots, S.length-1]$ 
in lines $12$ through $15$. Lines $16$ and $17$ insert a leaf node that indicates the end of the string 
in the tree.

We also implemented the query and deletion operations for the CloudPrefixTree.
Due to the limited space, we did not provide its pseudo code here.

\begin{algorithm}[t]
{\small
  \caption{$insert(n)$: Insert a key $n$ in the CloudBSTree.}
\label{algo:bstinsert}

\KwIn{The number $n$ to be inserted.} 
%\KwOut{All LR's that cover position $k$ or find no such LR.}

\smallskip 

\tcc{Start with root node, with id = $0$.}
$curNodeID \leftarrow 0$\;

\smallskip

\If{$The\ first\ key\ to\ insert$}{$Create\ the\ root\ node\ as\ a\ leaf\ and\ store\ it\ into\ cloud.$}
\Else{
%\tcc{Calculate $lr_1,lr_2,\ldots,lr_n$.}
\While{true}{
%$curChar \leftarrow S_i$\; %\tcp*{$\langle start, end\rangle$: start and ending position of $lr_k$.}

%\tcc{Whether in current node there is a child associated with character $S_i$.}
$curNode \leftarrow get\ tree\ node\ in\ cloud\ with\ curNodeID$\;
$curVal \leftarrow get\ the\ data\ item\ in\ curNode$\;

\smallskip  
%\tcc{Find the target in tree.}
\If(\tcp*[f]{Key already exists in the tree.}){$curVal == n$} {$return;$}

\smallskip
$idToGo \leftarrow followChild(curNode, n)$\;

  \If(\tcp*[f]{idToGo is not valid, find the place to insert.}){$idToGo < 0$}{$break$\;}
  \Else{$curNodeID \leftarrow idToGo;$}
}

\tcc{Create a child string. Child id and child data item are delimited with  $:$, then add it to current node.}
   $child \leftarrow ``nextNodeID : n$''\;
   $addChild(curNodeID, child)\; $\tcc{Update tree node in cloud via internet.}

\smallskip
$Create\ a\ new\ tree\ node\ for\ key\ n\ with\ id\ =\ nextNodeID$\;
$Add\ the\ new\ node\ into\ cloud\ DynamoDB\ table$\;
$nextNodeID \leftarrow nextNodeID + 1$\;
}%end else

}%end else
\end{algorithm}

\subsection{Cloud Binary Search Tree Implementation}
\label{bst}
In figure \ref{fig:bstSample}, we present a binary search tree that is maintained in cloud, named as
\emph{CloudBSTree}. The tree is created after inserting a collection of keys of integers.

We use the \emph{node\_id} and \emph{children\_set} in the same way as the CloudPrefixTree node.
%we use a numerical attribute \emph{node\_id} 
%to uniquely identify a tree node object in the DyanomaDB table in which the tree is stored.
%We use the same attribute \emph{children\_set} to explicitly store all references to the children of the current node,
%as well as the data item stored in each child. 
For example, 
the tree node with $node\_id = 0$ has two children represented with the \emph{children\_set} \{(``1:2''), (``5:7'')\}, with child ids \{1, 5\} and the data items stored in these children are
\{2, 7\} respectively. Note that an empty \emph{children\_set} value means a leaf node.

Unlike a CloudPrefixTree, we add a new attribute \emph{data\_item} for each tree node, 
to describe the data item stored in that tree node. The \emph{data\_item} attribute along with 
the string set \emph{children\_set} together to distinguish the left child and the right child within a node, 
since the \emph{children\_set} maintains no order itself. Also, this design allows to retrieve 
all children's information within a tree node in one cloud access, reducing the
number of network communications. 

\remove{
We present the query algorithm of a CloudBSTree in Algorithm \ref{algo:bstquery}.
Starting with the root node with id = $0$, we compare the key $n$ with the data item $curVal$ in the current node.
We stop the search if they are equal. Otherwise, we search in the left subtree of the current node if $n  <  curVal$, Or
we search in the right subtree of the current node if $n > curVal$. The function $followChild(curNode, n)$ at line $7$
returns the left child node id of the current node if $n < curVal$, or returns the right child node id if $n > curVal$, or returns
a negative number if $curNode$ is a leaf node.
}

Algorithm \ref{algo:bstinsert} describes the insertion operation of a CloudBSTree.
Starting with the root node, the while loop at lines $5$ to $14$ locates the place where
the new key will be inserted into the tree. 
The function $followChild(curNode, n)$ at line $10$
returns the left child node id of the current node if $n < curVal$, or returns the right child node id if $n > curVal$, or returns
a negative number if $curNode$ is a leaf node.
Once the place for insertion is located, at lines $15$ and $16$ we create and add a child string into 
the current node, which points to the new node for the key $n$ that will be inserted at lines $17$ and $18$.
Last, we increment the variable $nextNodeID$ for the next insertion.

We also implemented query and deletion operations for the CloudBSTree.
Due to the limited space, we did not provide its pseudo code here.

\begin{figure}[t]
\begin{center}
\includegraphics[scale=0.56]{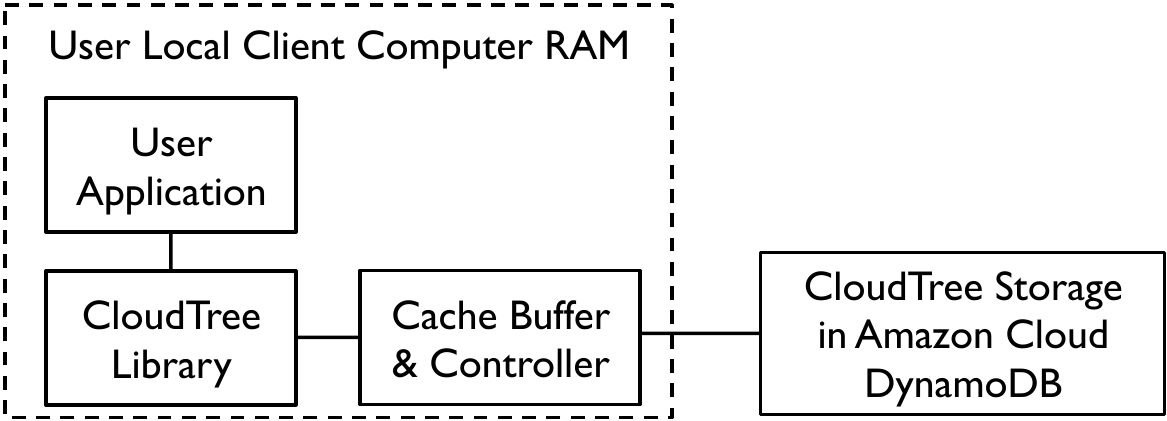}
\caption{CloudTree with a cache implementation in the RAM of user's client computer.}
\label{fig:cache}
\end{center}
%\vspace{-6 mm}  %originally not here
\end{figure}
%\vspace{-3 mm}  %original no here

\section{Optimizations}
\label{optimize}
We employ several optimization techniques to improve performance of the proposed CloudTree.
In particular, caching, prefetching, and aggregation are all effective in dramatically 
reducing the number of communications via Internet.

\subsection{Caching and Prefetching}
\label{cache}
The \emph{latency} cost we have to pay for each data access over the Internet could be very high, up to a hundred milliseconds. 
Therefore, when manipulating a CloudTree instance, it is crucial to reduce the number of data accesses via Internet
in order to achieve good performance.
Motivated by \emph{caching} techniques for reducing disk I/O operations, 
we add a cache module to the CloudTree.% --- a bounded cache buffer in the RAM of a user's client computer.

Figure \ref{fig:cache} illustrates the design of CloudTree with a cache --- a bounded cache buffer located in the RAM of a user's client computer and
a cache controller.
The cache is bridged right between the CloudTree library and CloudTree instances stored in a cloud. 
%The role of the cache is the 

The CloudTree cache here works in a similar way as a cache memory for a traditional hard drive.
When user applications access a tree node in a CloudTree instance, 
the request is first sent to the cache controller in figure \ref{fig:cache}.  
Then the cache controller checks whether the tree node has been cached in the local cache buffer.
If so, the copy of the tree node is retrieved from the cache. Otherwise, 
the cache controller retrieves from the cloud via Internet the requested tree node on user's behalf, 
then returns the tree node to the user, as well as saves it into the cache buffer.
When updating a tree node in a CloudTree, the cache controller updates the copy of the data first 
in the cache buffer if available, then updates the data item in cloud as well.

When the cache retrieves data from the cloud, we use \emph{prefetching}.
Particularly, when an application reads an uncached tree node with $node\_id = x$ --- 
in addition to the requested tree node $x$ --- the cache controller prefetches a collection of tree nodes
with node ids of $\{x + 1, x + 2, \ldots, x + prefetchSize\}$ and stores them into the cache buffer.
In this way, it is quite likely that subsequent tree node access requests can be answered with the data
in the cache, without communicating to the cloud via Internet for each access request.
We adopt the \emph{Least Recently Used (LRU)}  policy \cite{LRU12} to discard cached tree nodes when cache is full.

\subsection{Aggregations}
Each CloudTree operation (query, insert or delete etc.) may consist of multiple tree node insertions and updates.
For example, when inserting a string into a CloudPrefixTree, 
in a worst case scenario we have to create a tree node for each character in the string and insert into a underlying cloud database table.
In this case, we have to pay the Internet latency for \emph{each} tree node insertion, which is not efficient.

We aggregate these small operations or requests into one big request, then send the big request to the cloud.
In this way, we only pay the latency cost once for a group of data access or updates.  In particular,
When the cache controller performs prefetching, it fetches a collection of tree nodes in one transaction by using
underlying bulk cloud query operations. Similarly, when inserting or deleting a key from a cloud tree,
%for tree insertion and deletion operations, 
we aggregate multiple tree node insertions (or deletions) into one bulk insertion (or deletion) operation, thus reducing latency cost.
Note that we only aggregate tree node operations within the scope of each \emph{individual} CloudTree operation. In other words, the set of
tree node insertions for $tree.insert(keyA)$ and $tree.insert(keyB)$ would NOT be aggregated into one transaction.
The usefulness of these optimizations are evaluated in the next section.

%\section{Experiments} 
%dataset\\
%settings of tests, read capacity and write capacity\\
%client computer configuration\\
%purpose of the tests.\\
%network bandwidth and latency on the client side to the cloud\\
%
%\subsection{Test Results}
%
%\subsection{Analysis}
%the performance is determined by the number of access over the network.\\
%O(n) for trie, where n is the length of the prefix searched.\\
%Also, it is proportional to the the number of network access operations.
%O(logn), \\
%Unit time cost for database access, reading, update, and delete for pure dynamo data record on the client
\section{Experiments}
\label{tests}
We perform experiments with CloudTree instances stored in Amazon cloud. 
We verify the usefulness of the CloudTree library and 
measure time cost of insertion, query, and deletion for
two members of CloudTree, the CloudPrefixTree and the CloudBSTree, implemented 
with Java and the Amazon AWS Java SDK.

\begin{table}[t]
\begin{center}
\includegraphics[scale=0.48]{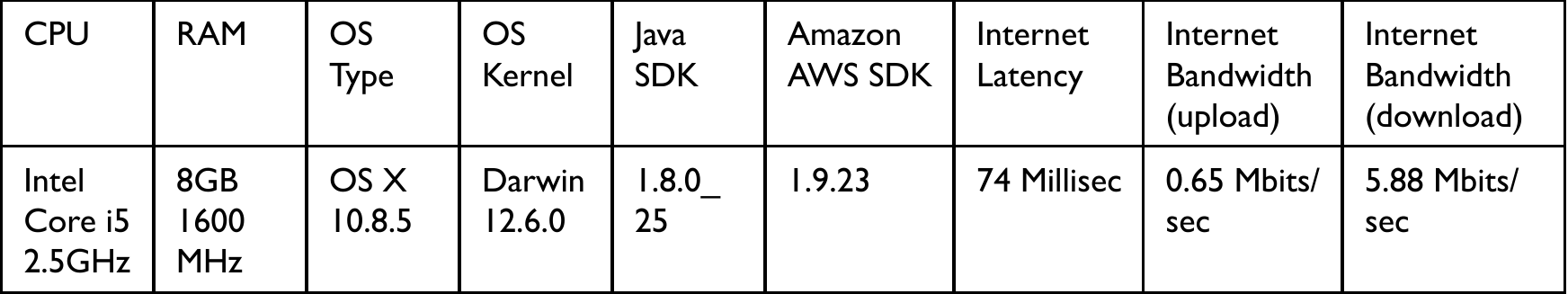}
\caption{Configuration parameters on the client computer. }
\label{fig:config}
\end{center}
%\vspace{-6 mm}  %originally not here
\end{table}

\begin{figure}[t]
\centering
\subfigure[Insertion]{
  \includegraphics[scale=0.38]{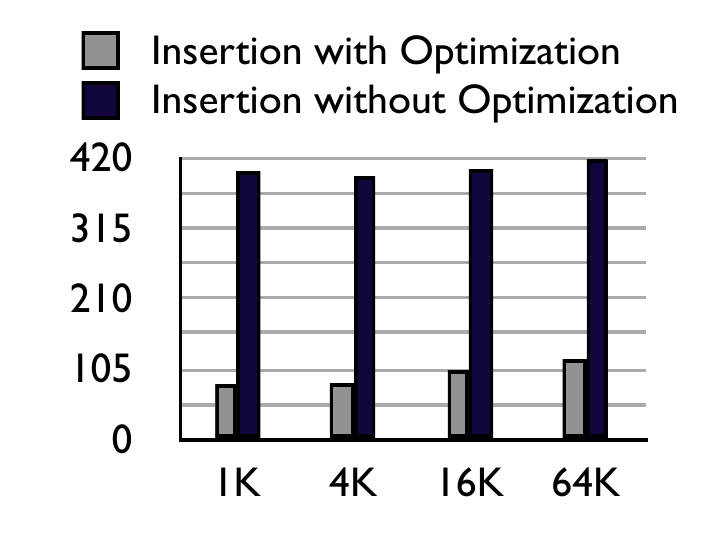}
   \label{fig:trie8in}
   }
   \hspace{-1\baselineskip}
\subfigure[Query]{
  \includegraphics[scale=0.38]{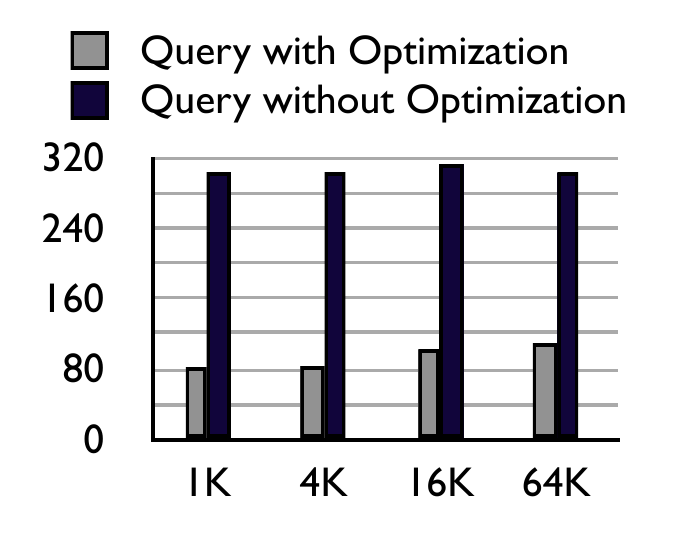}
   \label{fig:trie8qr}
   }
    \hspace{-1\baselineskip}
 \subfigure[Deletion]{
  \includegraphics[scale=0.38]{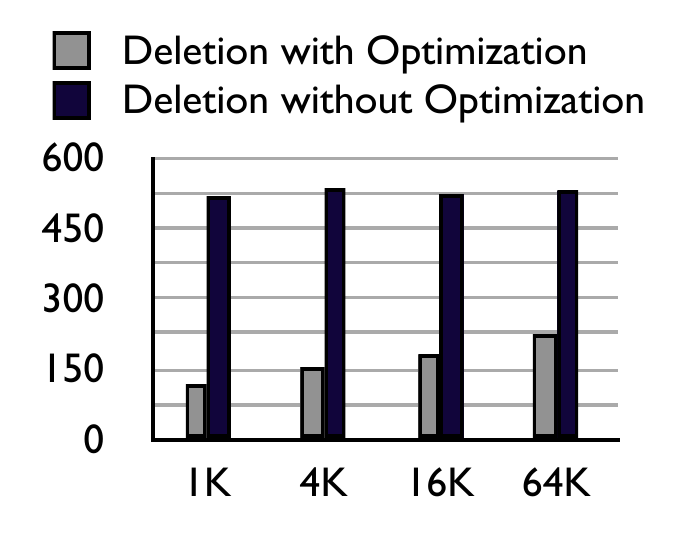}
   \label{fig:trie8dl}
   }
   \label{fig:trie8}
 \caption[Optional caption for list of figures]{CloudPrefixTree performance using strings of length eight.
  Horizontal axis denotes various number of strings maintained in the tree. Time cost shown on vertical axis is in milliseconds.
  %Time cost shown in the vertical axis  is in milliseconds.
 }
 %\vspace{-4 mm}  %originally not here
 \label{fig:trie8}
\end{figure}
%%%%%%%%

\begin{figure}[t]
\centering
\subfigure[4K dataset]{
  \includegraphics[scale=0.38]{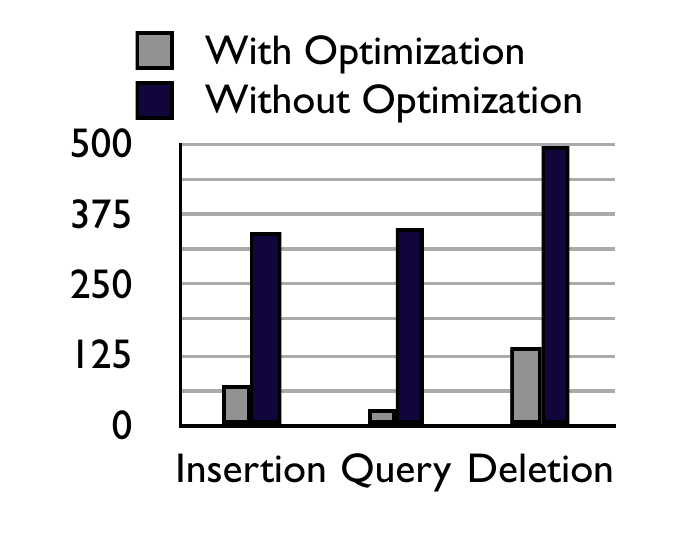}
   \label{fig:bst4kbal}
   }
   \hspace{-1\baselineskip}
\subfigure[8K dataset]{
  \includegraphics[scale=0.38]{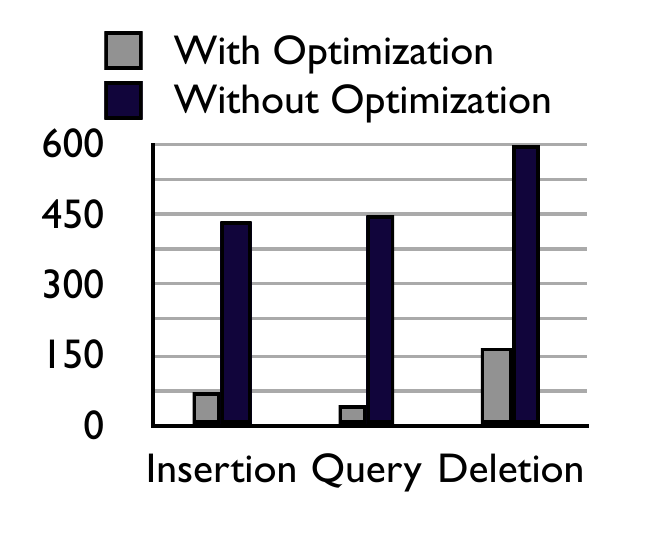}
   \label{fig:bst8kbal}
   }
    \hspace{-1\baselineskip}
 \subfigure[64K dataset]{
  \includegraphics[scale=0.38]{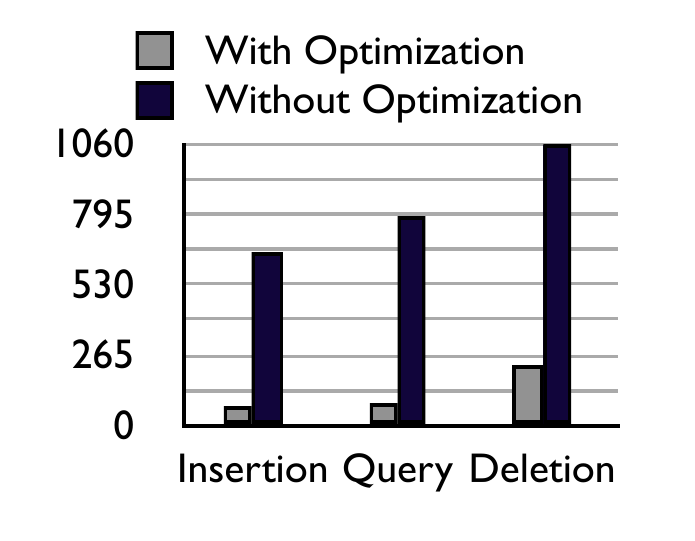}
   \label{fig:bst64kbal}
   }
   \label{fig:bst}
 \caption[Optional caption for list of figures]{CloudBSTree performance using datasets that result in balanced binary trees.
 Time cost shown on vertical axis is in milliseconds.
 }
 %\vspace{-4 mm}  %originally not here
 \label{fig:bst}
\end{figure}

%%%%%%%%%%%
\begin{figure}[t]
\centering
\subfigure[CloudPrefixTree]{
  \includegraphics[scale=0.37]{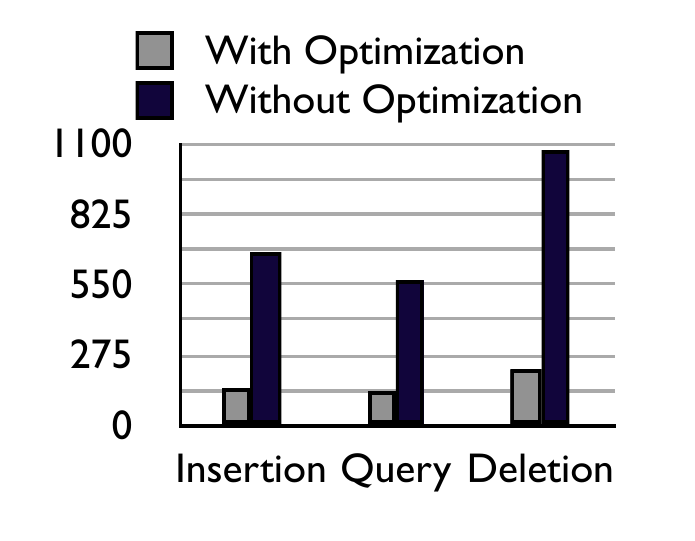}
   \label{fig:trie16s16k}
   }
   \hspace{-1\baselineskip}
\subfigure[CloudPrefixTree]{
  \includegraphics[scale=0.37]{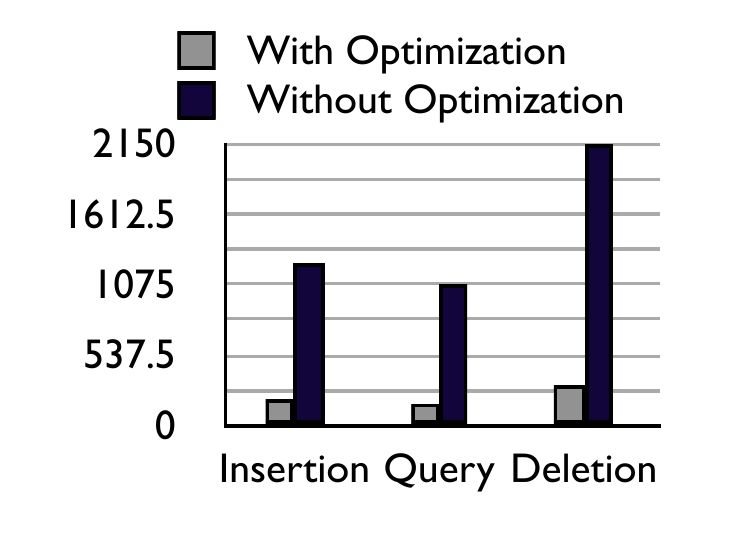}
   \label{fig:trie32s16k}
   }
    \hspace{-1\baselineskip}
 \subfigure[CloudBSTree]{
  \includegraphics[scale=0.37]{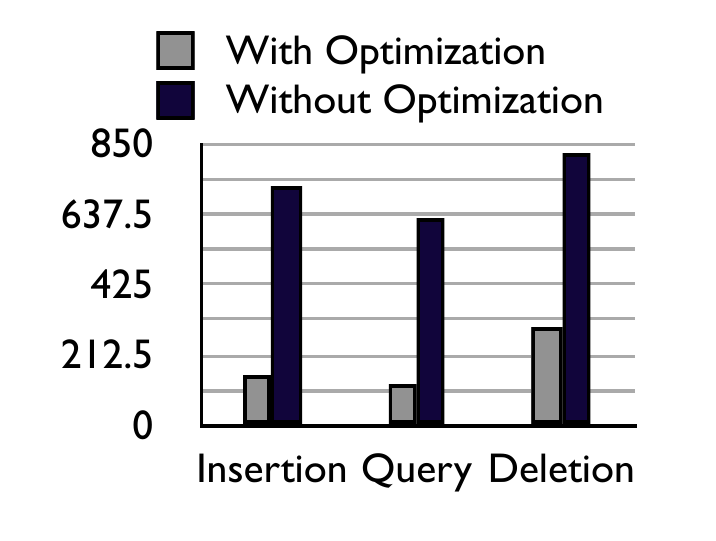}
   \label{fig:bst32krand}
   }
   %\label{fig:other1}
 \caption[Optional caption for list of figures]{Subfigure \subref{fig:trie16s16k} and \subref{fig:trie32s16k} show CloudPrefixTree performance using string length of 16 and 32 
 respectively. The total number of strings in tests is 16 thousand. Subfigure \subref{fig:bst32krand} illustrates CloudBSTree tests with a \emph{random} dataset of 32 thousand 
 unique integers. Time cost shown on vertical axis is in milliseconds.
 }
 %\vspace{-4 mm}  %originally not here
 \label{fig:other1}
\end{figure}
%%%%%%%%%%%%%%%%%%%
\begin{figure}[t]
\centering
\subfigure[CloudPrefixTree]{
  \includegraphics[scale=0.37]{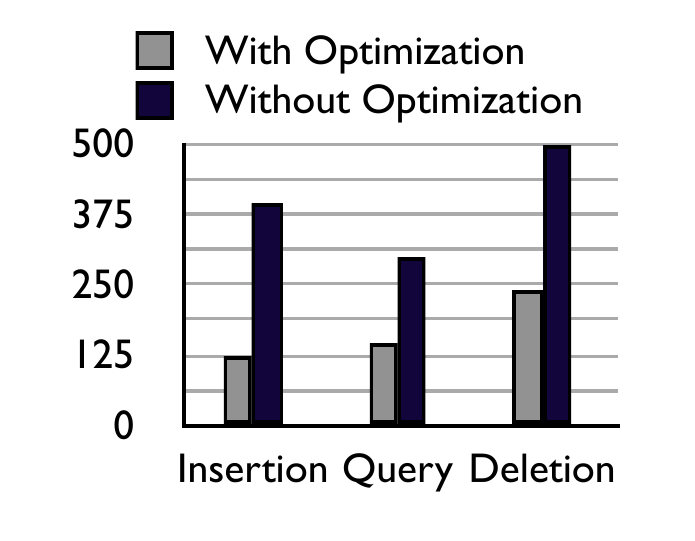}
   \label{fig:triep1c1}
   }
   \hspace{-1\baselineskip}
\subfigure[CloudBSTree]{
  \includegraphics[scale=0.37]{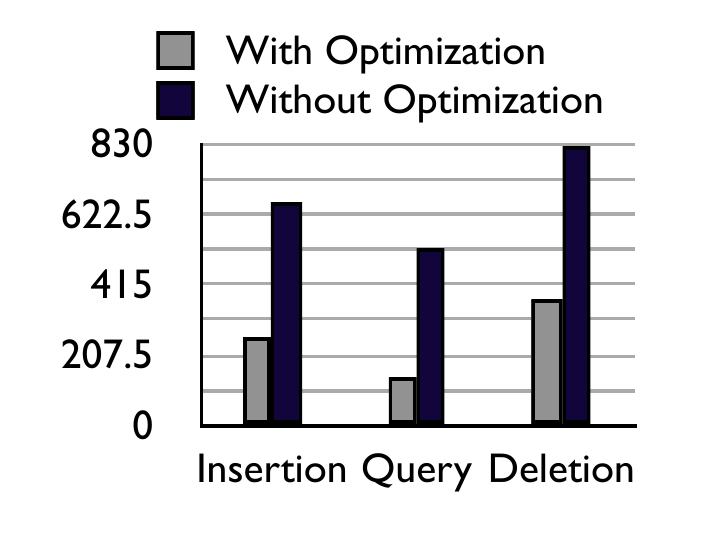}
   \label{fig:bstp1c1}
   }
%    \hspace{-1\baselineskip}
% \subfigure[CloudBSTree]{
%  \includegraphics[scale=0.37]{ctreepic/bst32kRand.pdf}
%   \label{fig:bst32krand}
%   }
%   \label{fig:other}
 \caption[Optional caption for list of figures]{Subfigure \subref{fig:triep1c1} and \subref{fig:bstp1c1} present performance of 
 a CloudPrefixTree and a CloudBSTree respectively. We use 16 thousand data items in the tests.
 In tests with optimizations, we use a small cache of only 1000 cache lines, but with prefetching \emph{disabled}. 
 Time cost shown on vertical axis is in milliseconds.
 }
 %\vspace{-4 mm}  %originally not here
 \label{fig:other2}
\end{figure}

\subsection{Experiment Setup}
\label{setup}

We initialize and manipulate a CloudTree instance on a client computer, whose configurations are shown in
table \ref{fig:config}. 
%We list the type of CPU, the RAM size, operating system(OS) and kernel version, Java SDK version,
%and the version of Amazon AWS SDK for Java. 
%We list the type of hardware and OS versions, as well as Java version and AWS SDK version.
We measure the last three parameters in table \ref{fig:config} with an cloud speed test service on CloudHarmony.com. 
$Internet\ Latency$ shows an average round trip time (RTT) between the client computer and an Amazon EC2 instance 
that is located in the same zone as the Amazon DynamoDB service that we use in the tests.
We also obtain bandwidth parameters between the Amazon cloud service and the client computer.

On the side of the cloud provider, we have to set up three parameters for each DynamoDB table.
%in which a CloudTree instance resides. 
%The provisioned throughput  \emph{readCapacity} and \emph{writeCapacity} is required
%for each DynamoDB table \cite{RWCapacity}. 
The throughput parameters specify the reserved throughput capacity for a created table  \cite{RWCapacity},
a positive number ranging from one to millions. Without specific notice, we use 500 for both read and write capacity.
The last parameter \emph{strongConsistency} describes whether we 
enforce retrieving the most recent value for a data item. %from a DynamoDB table. 
To maintain correctness and consistency of the data in a tree,
we always use the $strongConsistency = true$ option for all data retrieval.

With regard to test data, we download a 200MB text file of English literature from Pizza\&Chili~\footnotemark
\footnotetext{\url{http://pizzachili.dcc.uchile.cl/texts.html}}. %The size of the file is 200MB.
We divide the file into 8-character, 16-character or 32-character strings,
and use these strings for CloudPrefixTree tests. 

For CloudBSTree, 
we generate datasets in two ways. In the first approach, we create an array $A$ filled with contiguous integers ranging in $[0, H]$.
Then we re-order these numbers in $A$ into a list $L$ so that it results in a \emph{balanced} binary search tree after 
we insert items in $L$ in order.  
In the second approach,
after $A$ is created, we \emph{randomly} shuffle $A$ to generate an array $B$.
We insert elements in $B$ into CloudBSTree instances. %creating a tree with a \emph{random} layout.

When measuring performance, insertion time cost is an average time cost of the insertion for an entire dataset.
Query and deletion time cost is measured as an average using $200$ \emph{random} data items in the tree.
When performing tests with optimization enabled, if without special notice,
we use a value 25 for  the parameter $prefetchSize$ and a cache buffer with size of 10000 cache lines.
Each cache line is used to cache one tree node.
%Each cache line maintains an $\langle key, value \rangle$ pair,
%where the key is a tree node id and the value holds the entire node object associated with the tree node id.

\subsection{CloudPrefixTree Performance}
%
%figure 
In figure \ref{fig:trie8}, we show the performance of CloudPrefixTree.
We perform tests using various number of strings, with string length 8.
It takes roughly 100 milliseconds to insert and query a string of length 8 in the tree 
when optimizations are enabled, while it takes more than 300 milliseconds without optimizations. 
The proposed optimization techniques improve performance by a factor of 3.49 on average,
compared to an implementation without optimizations.

\paragraph{Observation}
%Two reasons accounts for the increases in the deletion time as the data size increases
%when optimization is enabled.
The performance with optimizations is quite similar for datasets of 1K and 4K,
while performance degrades as the data size increases from 16K to 64K;
this is especially noticeable for the deletion shown in figure \ref{fig:trie8}\subref{fig:trie8dl}.
However, the performance without optimizations is very stable.

\paragraph{Explanation}
While data size is less than 4K, most of the tree nodes in a tree can be cached, because
some strings can share a common prefix. While data size is quadrupled to 16K and further to 64K,
our cache starts to swap in and out tree nodes, incurring some overhead.

The aggregation can explain the performance degradation in the deletion shown in  
figure \ref{fig:trie8}\subref{fig:trie8dl}. As the data size becomes larger, it is more likely for multiple strings to
share a common prefix. That means, during string deletion, there are more tree node updates (removal of a child from a node)
 operations than tree node deletion operations 
 (a tree node contains only one child and removal of the child is equivalent to deleting the whole tree node). 
 With optimizations, we aggregate multiple node deletions into one bulk delete, but we do not aggregate multiple update operations due to the 
 lack of such mechanism in DynamoDB. 
 Due to possible data consistency errors during parallel reads, we do not represent a node update operation with a combination of
 a deletion and an insertion.
 Therefore, aggregation for node deletions is more aggressive when deleting a string from a small dataset than from a bigger dataset.
 On the other hand, \emph{without} optimizations, each tree node deletion or update incurs separate communication to the cloud.

 \paragraph{Other Tests}Figure \ref{fig:other1}\subref{fig:trie16s16k} and \ref{fig:other1}\subref{fig:trie32s16k} present the performance
 of CloudPrefixTree with strings of length 16 and 32.  We observe the consistent performance gains in these tests as we did with string length of 8.
 We also test the tree with a small cache of 1000 lines (with
 prefetching disabled) in figure \ref{fig:other2}\subref{fig:triep1c1}, in which caching and aggregation are still effective while performance gains
 degrade due to the small cache size.
 
 \paragraph{Analysis} In addition, we verify that CloudPrefixTree operations preserve the traditional complexity of $O(d)$, 
 where $d$ is the length of the string in the tree.
 Without optimizations, as we increase the string size from 8 to 16 and further to 32, we observe the time cost for 
 tree operation is increased linearly, proportional to the number of communications between the client and the cloud. 
 Under optimizations, as we increase the string size in the same way, the time cost for 
 tree operation is increased sublinearly, primarily because the use of caching, prefetching, and aggregation reduces communications to the cloud.

 %That is, the time cost for CloudPrefixTree operations
 %is proportional to the length of the string 
 
%First, when data size is less than 10K, all tree nodes can be certainly cached, 
%since we use cache size of 10000 lines. But when testing using 16K and 64K strings,
%the cache starts to swap in and out tree nodes, which incurs overheads.
%As we insert more strings that contains any characters, the average number of 
%children in each node grows. The number of bytes transferred via internet is increased
%for a group of tree nodes.Secondly, the linear search operation that check a character 
%Because of the aggregation. for 64 bigger dataset,
%more likely strings share common prefixes, more nodes updates less note deletion.
%Smaller dataset, less shared common prefix, (more deletion and less update)

\subsection{CloudBSTree Performance}
Figure \ref{fig:bst} describes the performance of CloudBSTree.
We observe that optimizations are effective in
reducing communications between client and the cloud, thus improving performance.
On average, a CloudBSTree with the optimizations is 5.5 times quicker than the 
implementation without the optimization.
As shown in figure \ref{fig:bst}\subref{fig:bst64kbal}, in the test with 64K data items, 
the performance is improved by a factor of 6.5 by using optimizations.

\paragraph{Observation}The insertion and 
query operations cost roughly 70 milliseconds, 
while deletion takes 228 milliseconds for the 64K dataset.
\paragraph{Explanation}
Queries, insertions, and deletions in a binary search tree logically have the same complexity of $O(log(n))$.
However, the deletion operation incurs more cloud communications. 
To delete an item $D$ in a tree, we first locate the tree node $Cur$ that contains $D$.
If $Cur$ is not a leaf, we have to find the largest item $ML$ in Cur's left subtree.
Next, we replace $D$ in the node $Cur$ with the value $ML$. At last, we delete $ML$ in the tree.
All updates and deletions mentioned above require cloud communications.

 In addition, we verify that the CloudBSTree preserves its traditional logarithmic nature.
As we increase the data size from 8K to 64K in figure \ref{fig:bst}, we observe the time cost for 
 tree operations is increased logarithmically. The total time cost is proportional to the 
 number of cloud communications. 
 Note that as the data size increases from 8K to 64K, the cache buffer starts to swap in and out 
 tree nodes, which incurs overhead.

\paragraph{Other Tests}
Figure \ref{fig:other1}\subref{fig:bst32krand} shows CloudBSTree performance using a \emph{random} dataset of 32K unique integers.
We also perform experiments with a \emph{random} dataset of one \emph{Million} unique integers. 
Under optimizations, query and insertion take 250 milliseconds on average and deletion takes
 630 milliseconds on average with the dataset of one Million random integers.
 We observe consistent performance gains with these datasets.
 In addition, in figure \ref{fig:other2}\subref{fig:bstp1c1}, we present the performance with a small cache of 1000 cache lines and with
  prefetching disabled. The proposed optimizations continue to be effective to reduce communication cost.

 %\subsection{Analysis}

%\section{Experiments} 
%dataset\\
%settings of tests, read capacity and write capacity\\
%client computer configuration\\
%purpose of the tests.\\
%network bandwidth and latency on the client side to the cloud\\
%
%\subsection{Test Results}
%
%\subsection{Analysis}
%the performance is determined by the number of access over the network.\\
%O(n) for trie, where n is the length of the prefix searched.\\
%Also, it is proportional to the the number of network access operations.
%O(logn), \\
%Unit time cost for database access, reading, update, and delete for pure dynamo data record on the client

\section{Conclusion and Future Work}
In this work, we propose a library \emph{CloudTree} that enables users on a client computer to organize big data into tree data
 structures of their choice that are physically stored in the cloud.
We use caching, prefetching and aggregation optimizations in the
  design and implementation of CloudTree to enhance performance.
  We have implemented the service of Binary Search Trees (BST) and
  Prefix Trees as current members in CloudTree library and have benchmarked
  their performance using the Amazon Cloud.
  
  In the future, we continue to improve the performance of CloudTree by using compression, 
  or different caching policies. For example, 
  we could load the top $n$ levels of a tree into the cache. 
  We will also be implementing the cloud B-tree and R-tree members for the CloudTree library.
\bibliographystyle{IEEEtran}
\bibliography{ctree}

% Generated by IEEEtran.bst, version: 1.13 (2008/09/30)
\begin{thebibliography}{10}
\providecommand{\url}[1]{#1}
\csname url@samestyle\endcsname
\providecommand{\newblock}{\relax}
\providecommand{\bibinfo}[2]{#2}
\providecommand{\BIBentrySTDinterwordspacing}{\spaceskip=0pt\relax}
\providecommand{\BIBentryALTinterwordstretchfactor}{4}
\providecommand{\BIBentryALTinterwordspacing}{\spaceskip=\fontdimen2\font plus
\BIBentryALTinterwordstretchfactor\fontdimen3\font minus
  \fontdimen4\font\relax}
\providecommand{\BIBforeignlanguage}[2]{{%
\expandafter\ifx\csname l@#1\endcsname\relax
\typeout{** WARNING: IEEEtran.bst: No hyphenation pattern has been}%
\typeout{** loaded for the language `#1'. Using the pattern for}%
\typeout{** the default language instead.}%
\else
\language=\csname l@#1\endcsname
\fi
#2}}
\providecommand{\BIBdecl}{\relax}
\BIBdecl

\bibitem{AV1987}
A.~Aggarwal and J.~S. Vitter, ``The {I/O} complexity of sorting and related
  problems (extended abstract),'' in \emph{Proceedings of the 14th
  International Colloquium on Automata, Languages and Programming (ICALP)},
  1987, pp. 467--478.

\bibitem{vitter:IO-book}
J.~S. Vitter, \emph{Algorithms and Data Structures for External Memory}, ser.
  Foundations and Trends in Theoretical Computer Science.\hskip 1em plus 0.5em
  minus 0.4em\relax Hanover, MA: Now Publishers, 2008.

\bibitem{FLPR1999}
M.~Frigo, C.~E. Leiserson, H.~Prokop, and S.~Ramachandran, ``Cache-oblivious
  algorithms,'' in \emph{Proceedings of the 40th Annual Symposium on
  Foundations of Computer Science (FOCS)}, 1999, pp. 285--298.

\bibitem{NM2007}
G.~Navarro and V.~M{\"{a}}kinen, ``Compressed full-text indexes,'' \emph{ACM
  Computing Surveys (CSUR)}, vol.~39, no.~1, 2007.

\bibitem{AW2004}
H.~Attiya and J.~Welch, Eds., \emph{Distributed Computing: Fundamentals,
  Simulations, and Advanced Topics}.\hskip 1em plus 0.5em minus 0.4em\relax
  Wiley-Interscience, 2004.

\bibitem{GKKG2004}
V.~K. A.~G. Ananth~Grama, George~Karypis, Ed., \emph{Introduction to Parallel
  Computing}.\hskip 1em plus 0.5em minus 0.4em\relax Addison-Wesley, 2004.

\bibitem{Papadopoulos11}
A.~Papadopoulos and D.~Katsaros, ``{A-Tree: Distributed Indexing of
  Multidimensional Data for Cloud Computing Environments},'' in
  \emph{Proceedings of the IEEE International Conference on Cloud Computing
  Technology and Science (CloudCom)}, 2011, pp. 407--414.

\bibitem{Dehne13}
F.~Dehne, Q.~Kong, a.~Rau-Chaplin, H.~Zaboli, and R.~Zhou, ``{A distributed
  tree data structure for real-time OLAP on cloud architectures},'' in
  \emph{Proceedings of the IEEE International Conference on Big Data}, 2013,
  pp. 499--505.

\bibitem{Yu14}
Y.~Yu, Y.~Zhu, W.~Ng, and J.~Samsudin, ``{An Efficient Multidimension Metadata
  Index and Search System for Cloud Data},'' in \emph{Proceedings of the IEEE
  International Conference on Cloud Computing Technology and Science
  (CloudCom)}, 2014, pp. 499--504.

\bibitem{Wu10}
S.~Wu, D.~Jiang, B.~B.~C. Ooi, and K.~K.-L. Wu, ``{Efficient B-tree Based
  Indexing for Cloud Data Processing},'' \emph{The Proceedings of the VLDB
  Endowment (PVLDB)}, vol.~3, no. 1-2, pp. 1207--1218, 2010.

\bibitem{Chang12}
C.~R. Chang, M.~J. Hsieh, J.~J. Wu, P.~Y. Wu, and P.~Liu, ``{HSQL: A highly
  scalable cloud database for multi-user query processing},'' \emph{Proceedings
  of the IEEE International Conference on Cloud Computing (CLOUD)}, pp.
  943--944, 2012.

\bibitem{Mo14}
Z.~Mo, Q.~Xiao, Y.~Zhou, and S.~Chen, ``{On Deletion of Outsourced Data in
  Cloud Computing},'' in \emph{Proceedings of the IEEE International Conference
  on Cloud Computing (CLOUD)}, 2014, pp. 344--351.

\bibitem{BM1970}
R.~Bayer and E.~McCreight, ``Organization and maintenance of large ordered
  indices,'' Boeing Scientific Research Labs, Mathematical and Information
  Sciences Report~20, 1970.

\bibitem{ferragina:string_B-tree}
P.~Ferragina and R.~Grossi, ``The {S}tring {B}-tree: {A} new data structure for
  string search in external memory and its applications,'' \emph{Journal of
  ACM}, vol.~46, no.~2, pp. 236--280, 1999.

\bibitem{bender:oblivious-string-B-tree}
M.~A. Bender, M.~Farach-Colton, and B.~Kuszmaul, ``Cache-oblivious string
  {B}-trees,'' in \emph{Proceedings of the ACM conference on Principles of
  Database Systems (PODS)}, 2006, pp. 233--242.

\bibitem{Behm11}
A.~Behm, C.~Li, and M.~J. Carey, ``Answering approximate string queries on
  large data sets using external memory,'' in \emph{Proceedings of the 2011
  IEEE International Conference on Data Engineering (ICDE)}, 2011, pp.
  888--899.

\bibitem{AmazonService13}
J.~Baron and S.~Kotecha, ``Storage options in the aws cloud,'' \emph{Amazon Web
  Services, Washington DC, Tech. Rep}, 2013.

\bibitem{LRU12}
G.~R. Mewhinney and M.~S. Srinivas, ``Optimized least recently used lookup
  cache,'' Sep. 2012, uS Patent 8,275,802.

\bibitem{RWCapacity}
\BIBentryALTinterwordspacing
Amazon, ``{Provisioned Throughput in Amazon DynamoDB},'' 2015. [Online].
  Available:
  \url{http://docs.aws.amazon.com/amazondynamodb/latest/developerguide/ProvisionedThroughputIntro.html}
\BIBentrySTDinterwordspacing

\end{thebibliography}

% that's all folks
\end{document}